\documentclass[12pt]{article}

\title{Vector-Meson-Dominance model contribution to $\pi^0 \to 4\gamma$}
\author{ Z.~K.~Silagadze\\
Budker Institute of Nuclear Physics,  630 090,
Novosibirsk, Russia
}
\date{}

\begin{document}

\maketitle

\begin{abstract}
Vector-Meson-Dominance model contribution to $\pi^0 \to 4\gamma$ is
calculated. The result confirms old estimates that this contribution is 
much smaller than the purely electromagnetic photon spliting graph
contribution calculated earlier.

\end{abstract}

In the framework of the Vector-Meson-Dominance (VMD) model, the decay 
$\pi^0\to 4\gamma$ was considered in \cite{V1} and \cite{V2}. However, 
no complete evaluation of the decay width can be found in these papers,
because the calculations are too tedious to be performed by hand. Only 
an upper limit $R\le (7\div 8.6)\times 10^{-16}$ was given in  \cite{V1}
as a result of partial evaluation of the squared decay amplitude. Here
$R$ stands for the ratio
$$R=\frac{\Gamma(\pi^0\to 4\gamma)}{\Gamma(\pi^0\to 2\gamma)}.$$
In \cite{V2} the completely different number $R\sim 10^{-9}$ is quoted, 
but we believe this result is erroneous.

The present experimental upper limit $2\times 10^{-8}$ on the 
$\pi^0\to 4\gamma$ decay branching ratio was obtained long ago \cite{VV3}.
Recently a Letter of Intent for a new experiment at PSI appeared \cite{VV4}
aimed to search $C$-noninvariant decay $\pi^0\to 3\gamma$. A by-product of
this experiment will be an improved measurement of the allowed decay 
$\pi^0\to 4\gamma$ as the most important background \cite{VV4}. 
Taking in mind this experimental situation, the complete calculation of the
VMD contribution seems desirable.

Using the advantage of the REDUCE Computer Algebra System \cite{V3}, we 
were able to perform this complete calculation and the results will be 
presented below.

\noindent For the standard $\pi^0\to 2\gamma$ amplitude 
($f_\pi\approx 93~\mathrm{MeV}$)
$$-i\frac{\alpha}{\pi f_\pi}\epsilon_{\mu\nu\lambda\sigma}\epsilon_1^\mu
\epsilon_2^\nu k_1^\lambda k_2^\sigma$$ 
the corresponding decay width looks like
$$\Gamma(\pi^0\to 2\gamma)=\frac{\alpha^2 m_\pi^3}{32\pi^3 f_\pi^2}\, .$$
While, by using the Kumar's parametrization of the covariant phase-space
\cite{V4}, the $\pi^0\to 4\gamma$ decay width can be written in the form
\cite{V4,V5}
$$\Gamma(\pi^0\to 4\gamma)=$$
$$=\frac{1}{4!}\,\frac{1}{2^{12}\pi^6 m_\pi^3}
\int\limits_{s_1^-}^{s_1^+} d s_1 \int\limits_{s_2^-}^{s_2^+} d s_2
\int\limits_{u_1^-}^{u_1^+} {d u_1\over \sqrt{\lambda(m_\pi^2,s_2,
s_2^\prime)}}\int\limits_{u_2^-}^{u_2^+} d u_2 \int\limits_{-1}^1
{d \zeta\over\sqrt{1-\zeta^2}}\;\overline{|M|^2}\, ,$$
where various Mandelstam-type variables are defined in Kumar's paper 
\cite{V4}.

\noindent In the VMD model, the $\pi^0\to 4\gamma$ decay amplitude has 
the form \cite{V1}
$$M=T(1,2;3,4)+T(2,1;3,4)+T(3,2;1,4)+T(2,3;1,4)+$$
$$+T(4,2;3,1)+T(2,4;3,1)+T(1,3;2,4)+T(3,1;2,4)+$$
$$+T(1,4;3,2)+T(4,1;3,2)+T(3,4;1,2)+T(4,3;1,2),$$
where for our choice of the coupling constants (note that $V\to\pi\gamma$
coupling constant is defined as $eg_{V\pi\gamma}$) and up to an irrelevant 
phase
$$T(1,2;3,4)=\frac{\alpha}{\pi f_\pi}\left [\frac{e^2g_{\rho\pi\gamma}^2}
{s_1-m_\rho^2}+\frac{e^2g_{\omega\pi\gamma}^2}{s_1-m_\omega^2}\right ]
\frac{1}{s_2-m_\pi^2}\times$$ $$\times
\epsilon_{\mu\nu\lambda\sigma}\epsilon_3^\mu k_3^\nu\epsilon_4^
\lambda k_4^\sigma\;\epsilon_{\alpha\beta\gamma\delta}(k_2+k_3+k_4)
^\beta\epsilon_1^\gamma k_1^\delta\; \epsilon_{\alpha\beta^\prime
\gamma^\prime\delta^\prime}(k_2+k_3+k_4)^{\beta^\prime}
\epsilon_2^{\gamma^\prime} k_2^{\delta^\prime}\, .$$
While calculating the total decay rate, the permutation symmetry of the 
phase space can be used to classify the 144 terms in the squared amplitude
into 7 types \cite{V1}:
\begin{itemize}
\item $|T(1,2;3,4)|^2$ -- 12 diagonal terms, $\;\;$ (i)
\item $T(1,2;3,4)T(2,1;3,4)$ -- 12 terms, $\;\;$ (ii)
\item $T(1,2;3,4)T(1,3;2,4)$ -- 24 terms, $\;\;$ (iii) 
\item $T(1,2;3,4)T(2,3;1,4)$ -- 24 terms, $\;\;$ (iv)
\item $T(1,2;3,4)T(3,1;2,4)$ -- 24 terms, $\;\;$ (v) 
\item $T(1,2;3,4)T(3,2;1,4)$ -- 24 terms, $\;\;$ (vi)
\item $T(1,2;3,4)T(3,4;1,2)$ -- 24 terms. $\;\;$ (vii)
\end{itemize}
All terms of a given symmetry type contribute equally in the total decay 
rate. Terms of the type (vii) turn out to be zero after doing the photons
polarization sums.

Introducing a dimensionless version of Kumar's invariant variables and using
the same notations for them:
$$s_1 = {1\over m_\pi^2} (q-k_1)^2,\; s_2 = {1\over m_\pi^2} (q-k_1-k_2)^2,$$
$$u_1 = {1\over m_\pi^2} (q-k_2)^2,\; u_2 = {1\over m_\pi^2} (q-k_3)^2,\; 
t_2={1\over m_\pi^2} (q-k_2-k_3)^2,$$
we get after performing the polarization sums by REDUCE
$$\frac{\Gamma(\pi^0\to 4\gamma)}{\Gamma(\pi^0\to 2\gamma)}=
\frac{\alpha}{4(4\pi)^4}\times$$ $$\times
\int\limits_0^1 d s_1 \int\limits_0^{s_1} d s_2
\int\limits_{s_2/s_1}^{1-s_1+s_2} {d u_1\over \sqrt{\lambda(1,s_2,s_2
^\prime)}}
\int\limits_{u_2^-}^{u_2^+} d u_2 \int\limits_{-1}^1 {d \zeta\over
\sqrt{1-\zeta^2}} \ F(s_1,s_2,u_1,u_2,t_2(\zeta)).$$
Note that the invariant variable $t_2$ is a linear function of the 
integration variable $\zeta$:
$$t_2=u_1-{1\over 2} (1+u_1)(1-u_2) -{1\over 2} (1-u_1)(1-u_2)
\left[ \xi \eta - \sqrt{(1-\xi^2)(1-\eta^2)} \zeta\right],$$
where
$$\xi = {\lambda(1,s_2,s_2^\prime) - (1-s_1)^2 + (1-u_1)^2 \over
2 (1-u_1) \sqrt{\lambda(1,s_2,s_2^\prime)} },$$
$$\eta = {(1-s_3^\prime)^2 - (1-u_2)^2 - \lambda(1,s_2,s_2^\prime) \over
2 (1-u_2) \sqrt{\lambda(1,s_2,s_2^\prime)} },$$
and $\lambda(x,y,z) = x^2+y^2+z^2-2(xy+xz+yz)$ is a conventional triangle
function. Besides
$$s_2^\prime = 1+s_2 -u_1 -s_1, \;\; s_3^\prime = 2 - s_1 - u_1 - u_2.$$
The limits of integration for the $u_2$-variable are
$$u_2^\pm = 1 - {1\over 2}(u_1+s_1) \pm {1\over 2}
\sqrt{\lambda(1,s_2,s_2^\prime)}.$$
The function $F(s_1,s_2,u_1,u_2,t_2)$ can be decomposed into six parts, each 
of them corresponding to the particular symmetry type mentioned above:
$$F=F_1+F_2+F_3+\frac{1}{2}\left [ F_4+F_5+F_6\right ].$$
Here
$$F_1=\left [\frac{a_\rho}
{s_1-b_\rho}+\frac{a_\omega}
{s_1-b_\omega}\right ]^2
\frac{1}{(s_2-1)^2}\;P_1, $$  
$$F_2=\left [\frac{a_\rho}
{s_1-b_\rho}+\frac{a_\omega}
{s_1-b_\omega}\right ]
\left [\frac{a_\rho}
{u_1-b_\rho}+\frac{a_\omega}
{u_1-b_\omega}\right ] \frac{1}{(s_2-1)^2}\;P_2,$$
$$F_3=\left [\frac{a_\rho}
{s_1-b_\rho}+\frac{a_\omega}
{s_1-b_\omega}\right ]^2\frac{1}{s_2-1}\;
\frac{1}{s_1-s_2+u_1+u_2-t_2-2}\; P_3,$$
$$F_4=\left [\frac{a_\rho}
{s_1-b_\rho}+\frac{a_\omega}
{s_1-b_\omega}\right ]
\left [\frac{a_\rho}
{u_1-b_\rho}+\frac{a_\omega}
{u_1-b_\omega}\right ]\frac{1}{s_2-1}\;\frac{1}{t_2-1}\;
P_4.$$
$$F_5=\left [\frac{a_\rho}
{s_1-b_\rho}+\frac{a_\omega}
{s_1-b_\omega}\right ]
\left [\frac{a_\rho}
{u_2-b_\rho}+\frac{a_\omega}
{u_2-b_\omega}\right ]\times $$ $$\times \frac{1}{s_2-1}\;
\frac{1}{s_1-s_2+u_1+u_2-t_2-2}\; P_5,$$
$$F_6=\left [\frac{a_\rho}
{s_1-b_\rho}+\frac{a_\omega}
{s_1-b_\omega}\right ]
\left [\frac{a_\rho}
{u_2-b_\rho}+\frac{a_\omega}
{u_2-b_\omega}\right ]\frac{1}{s_2-1}\;\frac{1}{t_2-1}\;
P_6,$$
where
$$\begin{array}{ll} 
a_\rho=(m_\pi g_{\rho\pi\gamma})^2,\,&
a_\omega=(m_\pi g_{\omega\pi\gamma})^2, \\
b_\rho=m_\rho^2/m_\pi^2,\, &
b_\omega=m_\omega^2/m_\pi^2,\end{array} $$
and $P_i,\,i=1\div 6$ are certain polynomials of the variables $s_1,s_2,u_1,
u_2,t_2$ given in the appendix.

For numerical calculations we need $g_{V\pi\gamma}$ coupling constants and
they can be estimated from the $V\to\pi^0\gamma$ decay widths with the 
following result \cite{V6}:
$$g_{\rho\pi\gamma}\approx 0.73~\mathrm{GeV}^{-1},\;\;
g_{\omega\pi\gamma}\approx 2.32~\mathrm{GeV}^{-1},$$
which translates into
$$a_\rho\approx 0.0097,\, a_\omega\approx 0.098,
\, b_\rho\approx 32.53,\,  b_\omega\approx 33.55.$$

\noindent After the numerical calculations, we obtain
$$R\approx 3.3\times 10^{-16}$$
in agreement with the estimates given in \cite{V1}. Therefore, the VMD 
model contribution in the $\pi^0\to 4\gamma$ decay width is indeed very 
small, many orders of magnitude smaller than the photon splitting graph
contribution $R\approx 2.6\times 10^{-11}$ \cite{V7}. 

Recently hadronic contribution to $\pi^0 \to 4\gamma$ was estimated by using
constituent quark loop model \cite{V8}. Earlier this contribution was 
calculated by using chiral perturbation theory  \cite{V9}. In both approaches 
the hadronic contribution was also found to be negligible.  

\section*{acknowledgments}
The author thanks E.A. Kuraev for bringing his attention to this problem.

\section*{Appendix}
The expressions for the $P_i(s_1,s_2,u_1,u_2,t_2),\,i=1\div 6,$ polynomials
are: 
$$P_1=s_2^2\left \{ \frac{}{}s_1^2[s_2^2+(1-s_1)^2+2(s_2-u_1)(1-s_1-u_1)]+
s_2(s_2-2u_1s_1)\right \},$$
$$P_2=s_2^2\left \{\frac{}{}s_2^2+s_1^2u_1(s_1-2s_2+3u_1-2)+s_1u_1[(1-u_1)^2
+s_2^2-2s_2u_1]\right \},$$
$$P_3=s_2^4(s_1^2+1)+s_2^3\left \{ \frac{}{} s_1^2(2-3u_1-u_2+2t_2)-2s_1^3+
s_1(u_2-u_1-2)+\right . $$ $$\left .+2(1-u_1-u_2+t_2)\frac{}{}\right\}+s_2^2
\left \{ \frac{}{}s_1^4+
s_1^3(4u_1+u_2-3t_2-2)+\right . $$ 
$$+s_1^2(4u_1^2+u_2^2+t_2^2-4u_1t_2+3u_1u_2-t_2u_2-2u_1-2u_2-t_2+2)+$$
$$+s_1(2u_1^2-u_2^2-2u_1t_2+u_1u_2+u_2t_2+3u_2-2t_2-2)+$$ $$\left . +
(1-u_1-u_2+t_2)^2\frac{}{} \right \}+s_2s_1\left \{ \frac{}{}s_1^3(t_2-u_1)
+\right . $$ $$+
s_1^2(5u_1t_2-4u_1^2-t_2^2-2u_1u_2-t_2u_2+u_1+2t_2)+ $$ $$
+s_1(1-u_1-u_2+t_2)(3u_1^2-u_1t_2+u_1u_2+u_1-3t_2)-$$ $$\left . 
-u_1(1-u_1-u_2+t_2)^2\frac{}{} \right \}+s_1^2(u_1^2+s_1u_1-s_1t_2-
u_1t_2+u_1u_2-u_1)^2,$$
$$P_4=s_2^2\left \{ \frac{}{}2t_2^2+u_1\left [ \frac{}{}(s_1+u_1)(u_1+u_2)^2
-2t_2(u_2+2u_1)-\right . \right. \hspace*{40mm} $$ $$ \left . \left . -
2s_1t_2(1+u_1+u_2-t_2)\frac{}{} \right ]\frac{}{} \right \}+
2s_2u_1\left \{ \frac{}{}s_1^2[t_2-u_1u_2
-(u_1-t_2)^2]+\right . $$ $$ +s_1t_2(u_2-2t_2-1)+
s_1u_1\left [ \frac{}{} u_1(3t_2-3u_2-2u_1+1)+u_2+2t_2-(u_2-t_2)^2\frac{}{} 
\right ]-$$
$$\left . - u_1(1-u_1-u_2+t_2)(2t_2-u_1^2-u_1u_2)\frac{}{} \right \}+$$ $$ +
u_1(s_1+u_1)\left [ \frac{}{}s_1t_2+u_1(1-u_1-u_2-s_1+t_2)\frac{}{} 
\right ]^2,$$
$$P_5=2s_2^4(s_1u_2+1)-2s_2^3\left \{ \frac{}{} 2s_1^2u_2+s_1(u_2^2+3u_1u_2
-2u_2t_2+u_1-3u_2+2)-\right .$$ $$\left . -2(1-u_1-u_2+t_2)\frac{}{} \right \}+
s_2^2\left \{ \frac{}{}2s_1^3u_2+\right .$$ $$+s_1^2(u_1^2+3u_2^2+8u_1u_2-
6u_2t_2+4u_1-6u_2-4t_2+2)+$$ $$ +s_1\left [ \frac{}{}u_2^3+2u_2^2(3u_1-t_2-2)+
u_2(7u_1^2-8u_1t_2-6u_1+2t_2^2+4t_2+8)+\right . $$ $$ \left . \left . +
4(u_1^2-u_1t_2-t_2-1)\frac{}{} \right ]+2(1-u_1-u_2+t_2)^2\frac{}{} \right \}
+$$ $$+2s_1s_2\left \{ \frac{}{}s_1^2[2t_2-u_1(1+u_1+u_2-t_2)]-s_1\left [ 
\frac{}{}u_1^3+u_1^2(1+4u_2-t_2)- \right . \right .$$ $$ \left .-
2u_1(1+u_2)(1-u_2+2t_2)+t_2(2+u_2t_2-2u_2+2t_2)\frac{}{} \right ]+$$ 
$$\left . +
(1-u_1-u_2+t_2)\left [ \frac{}{}u_1(2u_1u_2-u_2t_2+u_2^2+u_1-t_2-1)-
u_2t_2\frac{}{} \right ]\frac{}{} \right \}+ $$ $$ +
s_1(s_1+u_2)[u_1(1-u_1-u_2-s_1+t_2)+s_1t_2]^2,$$
$$P_6=s_2^2\left \{\frac{}{} s_1u_2[(u_1+u_2)^2-2t_2(1+u_1+u_2-t_2)]+2t_2^2
-2t_2u_1(2u_1+u_2)+ \right . $$ $$\left . \frac{}{} +
u_1^2(u_1+u_2)^2\right \}+2s_2\left \{ \frac{}{} u_2s_1^2(2u_1t_2-u_1u_2+
t_2-u_1^2-t_2^2)-\right . $$ $$-s_1\left [ \frac{}{} t_2^2(u_1u_2+u_1+u_2)- 
t_2(u_1^3+2u_1^2u_2+2u_1^2+2u_1u_2^2+u_2^2-u_2)+\right . $$ $$\left .  
\frac{}{} +u_1(u_1+u_2)(u_1^2+u_2^2+u_1u_2-u_2) \right ]+$$ $$\left . +
u_1^2(1-u_1-u_2+t_2)(u_1^2+u_1u_2-2t_2)\frac{}{}\right \}+ $$ $$+
(u_1^2+s_1u_2)[s_1(t_2-u_1)+u_1(1-u_1-u_2+t_2)]^2.$$

\end{document}